\documentclass[reprint,twocolumn,showpacs,nofootinbib,prd,amssymb]{revtex4}
\usepackage{latexsym}
\usepackage{graphicx,epsf, epsfig, amssymb, amsmath}

\usepackage[usenames,dvipsnames]{xcolor}

\usepackage{relsize}
\usepackage{dsfont}
\usepackage{todonotes}
\usepackage{textcomp}
\usepackage{float}
\usepackage{color}
\usepackage{mathrsfs}
\usepackage{soul}

\newcommand{\aap}{Astron.\ Astrophys.}
\newcommand{\mnras}{Mon.\ Not.\ R.\ Astron.\ Soc.}
\newcommand{\physrep}{Phys.\ Rep.}
\newcommand{\apjl}{Astrophys.\ J.\ Lett.}

\newcommand{\Msun}{$M_\odot$}
\newcommand{\Mmax}{M_*}
\newcommand{\Rmax}{R_*}
\newcommand{\csmax}{c_{sc*}}
\newcommand{\rhomax}{\rho_{c*}}
\newcommand{\Pmax}{P_{c*}}
\newcommand{\xmax}{x_*}
\newcommand{\ymax}{y_*}

\begin{document}

\title[Maximum-mass neutron stars]{Universal properties of maximum-mass neutron stars: a new tool to explore superdense matter}

\author{D.~D.~Ofengeim\footnote{ddofengeim@gmail.com}} 
\affiliation{Ioffe Institute, Polytekhnicheskaya 26, 194021 St. Petersburg, Russia}

\begin{abstract} 
I have demonstrated the existence of a tight correlation between the mass, radius, central density, and pressure of maximum-mass neutron stars modeled using diverse baryonic equations of state. A possible explanation for these correlations is provided. Simple analytic forms of such correlations are suggested and compared with observational constraints on the maximum mass of neutron stars and their radii. This gives a valuable tool to constrain maximum pressure and density that could be reached in stable neutron stars, assuming their equation of state is baryonic.
\end{abstract}

\date{\today}



\maketitle

\section{Introduction}

In spite of many efforts to explore the equation of state (EoS) and other properties of matter in the cores of neutron stars (NSs), the problem remains unsolved~\cite{LattPrak2016}, especially at densities $\rho\gtrsim 3\rho_0$, 
where $\rho_0 = 2.8\times 10^{14}\,$g$\,$cm$^{-3}$ is the density of the symmetric nuclear matter at saturation.

A relation between the EoS and observable properties of slowly-rotating NSs (considered here), such as the gravitational mass $M$ and circumferential radius $R$, can be calculated by solving the Tolman-Oppenheimer-Volkoff (TOV) equations~\cite{Tolman1939,OppVol1939}. The solution predicts the existence of the maximum mass $\Mmax$ of stable NSs. The corresponding radius $\Rmax$ is usually minimal for the NSs with a given EoS. The central density $\rhomax$ of the maximum-mass NS is the maximum density reachable in NS interiors. Due to the strong dengeneracy of the NS matter, many parameters $\xmax$ of the maximum-mass NSs, such as the central pressure $\Pmax$ and the central speed of sound $\csmax$, can be unambiguously expressed through $\rhomax$, $\xmax=\xmax(\rhomax)$. Another property is that the Oppenheimer-Volkoff (OV) map [from $P(\rho)$ to $R(M)$ using the TOV equations] is a bijection~\cite{Lind1992}: for a given $M-R$ relation, one can unambiguously derive a central density $\rho_{c}$ and pressure $P_{c}$ for each NS with a mass $M$ and thus, reconstruct $P(\rho)$. 

There are several theoretical constraints on the properties of the maximum-mass NSs (see Ref.~\cite{LattPrak2016} for a comprehensive review). One of them follows from the causality condition $\csmax < c$ ($c$ is the speed of light), which is necessary but not sufficient. First, $c_s$ may be non-monotonic with $\rho$ (although typically $\csmax$ is the maximum $c_s$ at $\rho\leqslant\rhomax$). Second, strictly speaking, there should be $c_s<c$ at any $\rho$, both greater and less than $\rhomax$. However, it can be argued (e.g., Refs.~\cite{Read+2009,HPY2007}) that the validity of an EoS model should be evaluated only at $\rho\leqslant\rhomax$.

The most reliable observational constraints on the properties of the maximum-mass NSs come from accurate mass determinations of massive pulsars in compact binaries, e.g. the PSR~J0348+0432 with $M = 2.01\pm 0.04\,$\Msun~\cite{Ant2013} and the PSR~J0740+6620 with $M=2.14_{-0.09}^{+0.10}\,$\Msun~\cite{Crom2019}. These values should place a lower limit on $\Mmax$. An opposite restriction comes from the GW170817 event~\cite{GW170817}, observed as a gravitational-wave signal from a NS-NS merger, and its electromagnetic counterpart. Simultaneous analysis of these data gives~\cite{Rezz2018} an upper limit, $\Mmax < 2.16_{-0.15}^{+0.17}\,$\Msun. It is currently reasonable to constrain the NS maximum mass as $1.97\,M_\odot < \Mmax < 2.33\, M_\odot$ and treat the EoS models with $\Mmax<1.97\,M_\odot$ as unrealistic.

There exist many diverse NS EoS models (see, e.g., Refs.~\cite{HPY2007,BurgFant2018}). However, Lindblom~\cite{Lind2010} developed a spectral representation for physically motivated $\rho(P)$ functions. He showed that, for a wide class of EoS models, a few terms of the spectral expansion are sufficient to approximate accurately the $P-\rho$ dependence in the NS core. In particular, an approximation by first two terms has $3-8\%$ root mean square (rms) relative error. The set of 34 EoSs considered in Ref.~\cite{Lind2010} contains nucleon, hyperon, meson-condensate, quark, and hybrid models. A maximum relative error of about $20\%$ emerges for the models with a strong phase transition in the NS core. 

In other words, for many diverse EoSs, to a good approximation their $P-\rho$ relations belong to a single 2-parameter family. This means, owing to the bijectivity of the OV map, that for a wide class of EoS models, the sets of points $(M,R,\rho_c, P_c,...)$ approximately form a family of 2-parametric multidimensional curves. Therefore, for each $x=\rho_c,P_c,\ldots$ the 3D-points $(\Mmax,\Rmax,\xmax)$ approximately form a single surface. 

In this work, I have shown that for a wide class of baryonic EoSs (wider than in Ref.\ \cite{Lind2010}) such surfaces indeed exist and can be fitted by analytic expressions. Combining these fits with several constraints on $\Mmax$, $\Rmax$, and their relations to $\xmax$, one can obtain novel restrictions on the properties of superdense matter in the maximum-mass NSs.

\section{Analytic fits}

\begin{figure*}
\includegraphics[width=0.85\textwidth]{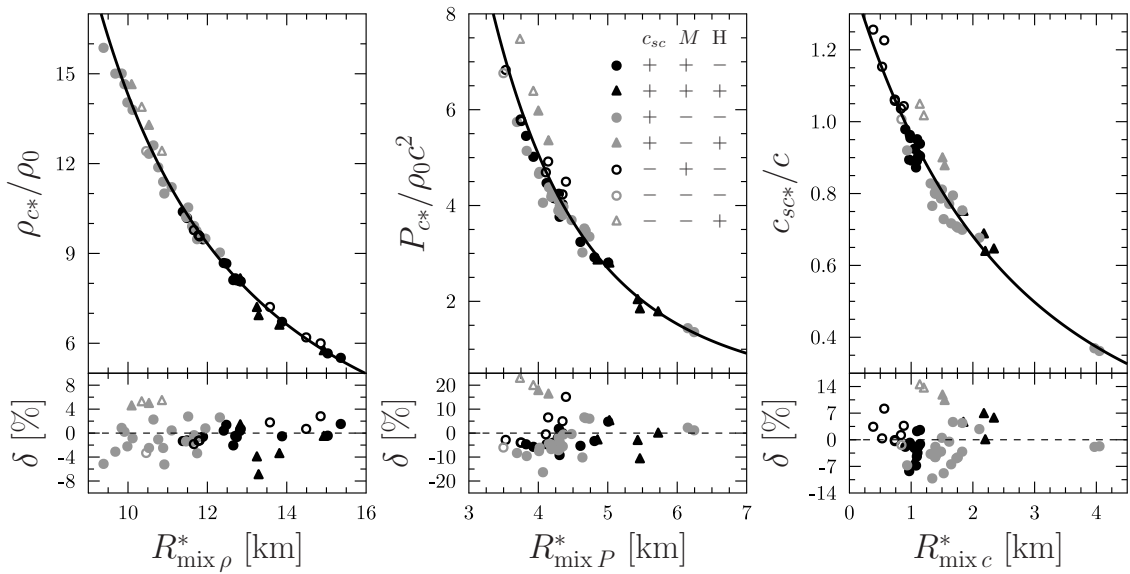}
\caption{\label{fig:fit_MR} Fitted correlations between $\Mmax$, $\Rmax$, and $\rhomax$ (left), $\Pmax$ (middle) and $\csmax$ (right) described by Eq.~(\ref{eq:x-MR}) (solid lines). Here $R^*_{{\rm mix}\, x} = \Rmax\cos\varphi_x + r_{g*}\sin\varphi_x$ for $\xmax=\rhomax,\Pmax,\csmax$, and $\varphi_x$ is given in Table~\ref{tab:x-MR} for each $\xmax$. Symbols refer to selected EoSs. Filled/open symbols show subluminal/superluminal models (+/-- signs in the ``$c_{sc}$'' column in the legend). Black/grey symbols mark the EoSs that obey/disobey the condition $\Mmax\geqslant 1.97\,$\Msun (+/-- in the ``$M$'' column). Circles/triangles correspond to the nucleon/hyperon EoSs (--/+ in the ``H'' column). Bottom plots show relative fit errors.}
\end{figure*}

Instead of using a specific 2-parameter family of EoSs, I have collected 50 EoS models of superdense nuclear matter based on a diverse microphysics input. If the desired correlations are found, it would mean that $P(\rho)$ relations for this EoS set can be viewed as a single 2-parametric function family.

The collected set contains: 17 PAL- \cite{PAL1988} and PAPAL-like \cite{PAPAL1992} EoSs, used  \cite{Yak2011} for studying NS cooling; BPAL12, BGN1,2, BGN1,2H1,2, and BBB1,2 models (e.g., Ref.\ \cite{HPY2007}, Chap.~6); the APR EoS \cite{APR1998} and 4 its HHJ-like \cite{HHJ2000} modifications, APR~I--IV \cite{Gus2005,KKPY2014}; the SLy4 EoS \cite{DH2001}; BSk19--22,24--26 EoSs \cite{BSk2013,BSk2018}; BHF EoS from Ref.~\cite{SBH2018}; the NL3, NL3$\omega\rho$, and DDME2 models from Ref.~\cite{Fortin2016}; the GM1A, GM1'B, and TM1C EoSs \cite{GHK2014}; hyperonic NL3$\omega\rho$ EoS \cite{Horowitz01}; the FSU2H EoS \cite{Prov+2018}; and, finally, 2 models of free neutron and neutron-proton-electron gases (e.g. Ref.~\cite{ShapTeuk1983}), glued to the BSk24 crust near the neutron-drip point. 

All these EoSs have a crust, models for which are also different. Details of crust models and crust-core matching are discussed in the references on specific EoSs cited above but seem insignificant for the maximum-mass NSs, which have a very thin crust. To emphasize the universality of the correlations (see below), both old and modern models are included in the EoS set. Some of the modern EoSs, like BSk24, BHF, and FSU2H, are consistent with the constraints inferred from a simultaneous analysis of nuclear-physical and astrophysical data (e.g, with the results of Refs.~\cite{Tews+2018,Greif+2019}). Among the 50 selected EoSs, 40 are subluminal (obey the causality condition), $\csmax<c$, while the others are superluminal (acasual); 24 models have realistically high maximum masses, $\Mmax>1.97\,$\Msun, the rest of them do not; 9 of them have hyperons in the NS cores, the remaining 41 are nucleonic. None of them allows for a phase transition to exotic matter (a meson condensate or deconfined quarks). Within this EoS set, $\Mmax$ varies from $\sim 0.7$ to $\sim 2.8\,$\Msun, $\Rmax$ ranges from $\sim 8.5$ to $\sim 13.5\,$km and $\csmax/c$ from 0.35 to 1.25. The set contains the EoSs that satisfy and disobey realistic criteria of causality and having a high enough $\Mmax$. This is necessary since I will substitute the fits into the equations like $\csmax(\Mmax,\Rmax)=c$, etc., and prefer to make the edges of the fitting domain to be distant from a solution of these equations.

\begin{table}
\centering
\caption{\label{tab:x-MR}  Fit parameters in Eq.~(\ref{eq:x-MR}), rms and maximum relative fit errors.}
\renewcommand{\arraystretch}{1.4}
\setlength{\tabcolsep}{0.14cm}
\begin{tabular}{l|c|cccc|cc}
\hline\hline
$\xmax$        & $x_0$        & $a_x\,$[km]  & $\varphi_x$  & $b_x\,$[km]  & $p_x$    & rms      & max  \\
\hline
$\rhomax$      & $\rho_0$     & $33.30$      & $0.364$      & $2.91$       & $1.72$   & $0.026$  & $0.071$      \\
$\Pmax$        & $\rho_0 c^2$ & $10.72$      & $-0.693$     & $-3.90$      & $5.32$   & $0.081$  & $0.23$     \\
$\csmax$       & $c$          & $9.198$      & $-0.949$     & $-8.30$      & $3.39$   & $0.055$  & $0.14$     \\
\hline\hline
\end{tabular}
\end{table}

\begin{figure*}
\includegraphics[width=0.85\textwidth]{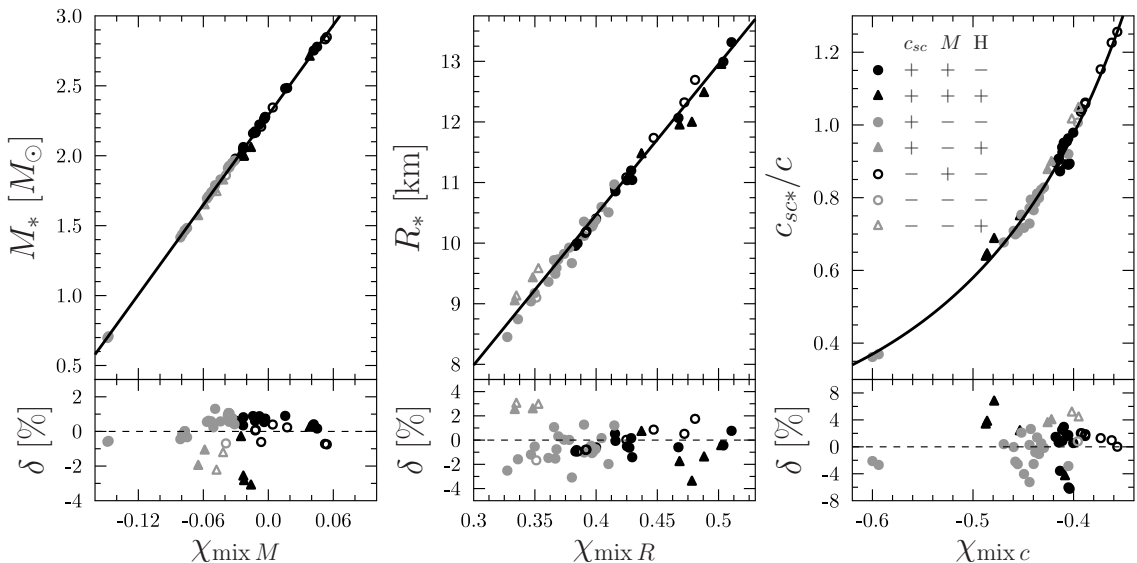}
\caption{\label{fig:fit_Prho} Correlations between $\Pmax$, $\rhomax$ and $\Mmax$ (left), $\Rmax$ (middle), and $\csmax$ (right) described by Eq.~(\ref{eq:y-Prho}) (solid lines). Here $\chi_{{\rm mix}\, y} = (\rho_0/\rhomax)^{1/1.72}\cos\Phi_y + (\rho_0 c^2/\Pmax)^{1/5.32}\sin\Phi_y$ for $\ymax=\Mmax,\Rmax,\csmax$, and $\Phi_y$ is given in Table~\ref{tab:y-Prho} for each $\ymax$. Symbols refer to selected EoSs with the same legend as in Fig.~\ref{fig:fit_MR}.}
\end{figure*}

The correlations in question $\xmax(\Mmax,\Rmax)$, with $\xmax=\rhomax,\Pmax,\csmax$, can be fitted by
\begin{equation}
\label{eq:x-MR}
\xmax = x_0 \left( \frac{a_x}{\Rmax \cos\varphi_x + r_{g*}\sin\varphi_x - b_x} \right)^{p_x}.
\end{equation}
Here, $r_{g*} = 2G \Mmax/c^2$ ($G$ is the gravitational constant), and the parameters $x_0$, $p_x$, $a_x$, $\varphi_x$, $b_x$, as well as their rms and maximum relative fit errors are listed in Table~\ref{tab:x-MR}. The dimensional constant $x_0$ has been fixed; $a_x$, $b_x$, $\varphi_x$, and $p_x$ have been varied to minimize the rms. The ``mixing angle'' $\varphi_x$ has geometric interpretation: looking at a 3D plot of the set of points $(\Rmax,\Mmax,\xmax)$, one can discover a plane, parallel to the $\xmax$ axis; projections of all points to this plane approximately form a single line. Actually, this plane is shown in Fig.~\ref{fig:fit_MR} for the three $\xmax$ types; $\varphi_x$ is the angle between this plane and the $\Rmax$ axis.

Fig.~\ref{fig:fit_MR} makes the correlations apparent and shows how accurate the fits are. The smallest errors occur for $\rhomax$, the largest ones are for $\Pmax$. The maximum errors always occur for hyperon EoSs (the triangles in Fig.~\ref{fig:fit_MR}). Anyway, the correlations $\rhomax(\Mmax,\Rmax)$ and $\Pmax(\Mmax,\Rmax)$ are tight enough to suggest that the $P-\rho$ relations of the 50 selected EoSs do belong to some 2-parameter family (the EoSs are typically the worst to fit). The EoS set used here partially overlaps with the collection from Lindblom's work~\cite{Lind2010}. Thus hereafter I refer to this 2-parameter family as to the Lindblom family. Since $c_s(\rho)$ is determined by the derivative ${\rm d}P(\rho)/{\rm d}\rho$, the $c_s(\rho)$ functions might form a 2-parameter family with worse accuracy. The unexpectedly high precision of the $\csmax(\Mmax,\Rmax)$ fit is probably because all the 50 EoSs are baryonic, without exotic matter phases and strong phase transitions. If the real EoS of the NS core possesses a transition to some exotic phase at $\rho<\rhomax$, its $\csmax$ may deviate stronger from Eq.~(\ref{eq:x-MR}).

\begin{table}
\centering
\caption{\label{tab:y-Prho} Parameters in Eq.~(\ref{eq:y-Prho}), rms and maximum relative deviations.}
\renewcommand{\arraystretch}{1.4}
\setlength{\tabcolsep}{0.12cm}
\begin{tabular}{l|c|cccc|cc}
\hline\hline
$y_*$        & $y_0$   & $A_y$    & $\Phi_y$ & $B_y$     & $q_y$    & rms      & max  \\
\hline
$\Mmax$      & \Msun   & $10.71$  & $-0.373$ & $-0.214$  & $1$      & $0.010$  & $0.031$ \\
$\Rmax$      & $1\,$km & $24.82$  & $0.178$  & $-0.0218$ & $1$      & $0.014$  & $0.034$ \\
$\csmax$     & $c$     & $-1.668$ & $-0.888$ & $0.204$   & $-3.39$  & $0.028$  & $0.067$ \\
\hline\hline
\end{tabular}
\end{table}

The relations~(\ref{eq:x-MR}) can be inverted. Namely, let us consider the fits $\rhomax(\Mmax,\Rmax)$ and $\Pmax(\Mmax,\Rmax)$ as a system of equations for $M_*$ and $R_*$, and derive formulas  for the correlations $\Mmax(\Pmax,\rhomax)$ and $\Rmax(\Pmax,\rhomax)$. Substituting them into Eq.~(\ref{eq:x-MR}) for $\csmax$, one can obtain the formula for the correlation between the speed of sound, the pressure and the density in the centers of the maximum-mass NSs. The resulting formulae for the $\ymax(\Pmax,\rhomax)$ correlations, with $\ymax = \Mmax,\Rmax,\csmax$, are
\begin{multline}
\label{eq:y-Prho}
\ymax={y_0}  \left\{ A_y \left[ \left( {\rho_0}/{\rhomax} \right)^{1/1.72}\cos\Phi_y \right.\right.
\\
+ \left.\left. \left( {\rho_0 c^2}/{\Pmax} \right)^{1/5.32}\sin\Phi_y - B_y \right] \right\}^{q_y}.
\end{multline}
The dimensional parameters $y_0$ are introduced for convenience; other parameters are derived from Eq.~(\ref{eq:x-MR}); all of them are listed in Table~\ref{tab:y-Prho}. These expressions are even more accurate than Eq.~(\ref{eq:x-MR}). This is also shown in Fig.~\ref{fig:fit_Prho}. Again, the maximum errors occur for hyperon EoSs.

\section{Constraining properties of maximum-mass neutron stars}

\begin{figure*}
\centering
\includegraphics[width=0.9\textwidth]{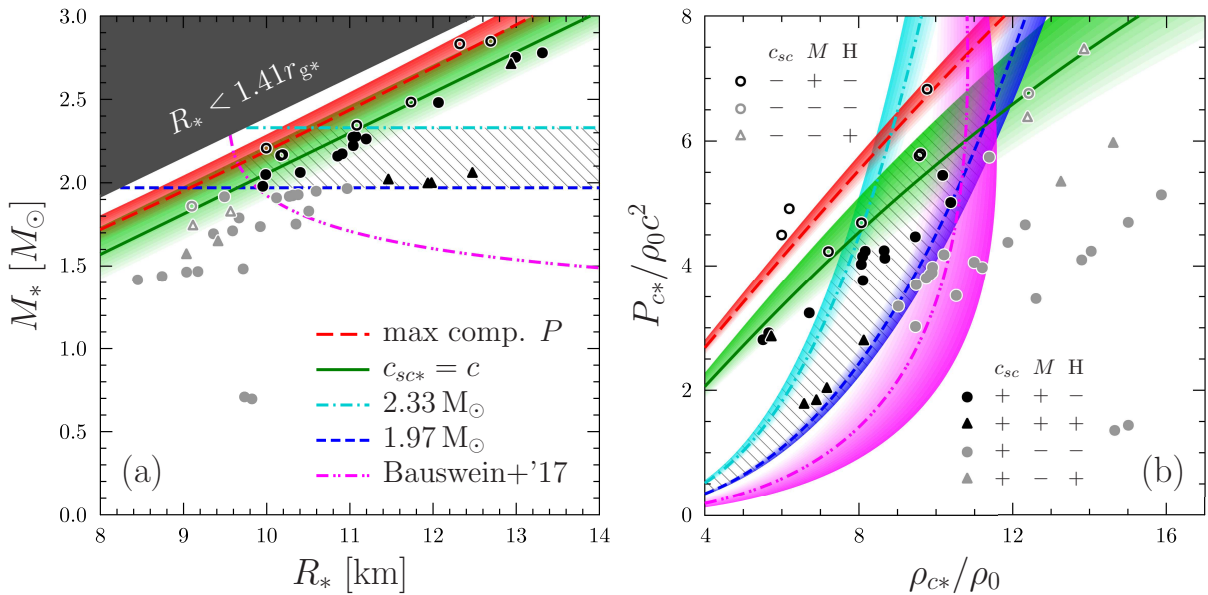}
\caption{\label{fig:MRPrho} 
Constraining properties of the maximum-mass NSs on the (a) $\Mmax-\Rmax$ and (b) $\Pmax-\rhomax$ planes. Symbols mark the $(\Mmax,\Rmax)$ and $(\Pmax,\rhomax)$ points for the EoSs used to calibrate the fits~(\ref{eq:x-MR}) and (\ref{eq:y-Prho}); the notations are the same as in Fig.~\ref{fig:fit_MR}. The darkened domain in panel~(a) is excluded directly due to maximum compactness. The solid lines show the upper limit of the causality constraint [Eq.~(\ref{eq:lumin-Rrg}) in (a) and Eq.~(\ref{eq:lumin-Prho}) in (b)]. The long-dashed lines are for the upper limits of the maximum compactness condition combined with the fits (\ref{eq:x-MR}) and (\ref{eq:y-Prho}) [Eq.~(\ref{eq:maxcomp-MR}) in (a) and Eq.~(\ref{eq:maxcomp-Prho}) in (b)]. The short-dashed and dot-dashed lines show (a) the boundaries of the observational constraints on $\Mmax$ and (b) their mapping by Eq.~(\ref{eq:Mfixed-Prho}). The double-dot-dashed lines are the lower limits of the constraint~(\ref{eq:Bauswein}) by Bauswein et al., Ref.~\cite{Baus2017}. For each line, the surrounding shaded areas show estimated fit uncertainties. For each strip, the shading transparency decreases towards an area forbidden by a corresponding condition. The hatched area is the domain allowed by all the constraints applied together. See text for details.}
\end{figure*}

In Fig.~\ref{fig:MRPrho}, symbols mark the selected EoSs in the $\Mmax-\Rmax$ and $\Pmax-\rhomax$ planes. One can see that the models which simultaneously satisfy the causality condition $\csmax < c$ and the observational constraint $\Mmax>1.97\,$\Msun\ (filled black symbols) form a separated domain in each plane. The boundaries of these domains can be derived analytically using Eqs.~(\ref{eq:x-MR}) and~(\ref{eq:y-Prho}), as well as some other constraints.

To map the upper boundary of the causality condition to the $\Mmax-\Rmax$ plane, one has to solve the equation $\csmax(\Mmax,\Rmax)\times (1\pm 0.14) = c$, where Eq.~(\ref{eq:x-MR}) for $\xmax=\csmax$ is used, and the factor $1\pm 0.14$ estimates the fit error. The solution is 
\begin{equation}
\label{eq:lumin-Rrg}
\Mmax/\text{\Msun} - 0.243 \Rmax /(1\text{km}) \approx -0.37 \pm 0.16.
\end{equation}
Similarly, to get the upper limit of the causality condition in the $\Pmax-\rhomax$ plane, one should substitute Eq.~(\ref{eq:y-Prho}) into the equation $\csmax(\Pmax,\rhomax)\times (1\pm 0.07) = c$ (with a 7\% fit error). Its solution is
\begin{equation}
\label{eq:lumin-Prho}
\left( \frac{\rho_0 c^2}{\Pmax} \right)^{1/5.32} -0.813 \left( \frac{\rho_0}{\rhomax} \right)^{1/1.72} \approx 0.510 \pm 0.016.
\end{equation}
The mean values of the upper limits~(\ref{eq:lumin-Rrg}) and~(\ref{eq:lumin-Prho}) are plotted in Fig.~\ref{fig:MRPrho} by solid lines. Shaded semi-transparent strips around these lines reflect fit uncertainties. The shading transparency decreases towards the acasual area. 

If one does not rely on the $\csmax$ fits, a more reliable (but less stringent) constraint can be derived using fit~(\ref{eq:x-MR}) for $\Pmax$ and fit~(\ref{eq:y-Prho}) for $\Mmax$. One can apply these fits to the restrictions imposed by the maximally-compact EoS (e.g. Sec. 2.2 of Ref.~\cite{LattPrak2016}). This is the EoS with a zero pressure below some threshold density $\rho_{\rm th}$, and $P=(\rho-\rho_{\rm th}) c^2$ at higher $\rho$. It yields \cite{LattPrak2016} upper limits of $\Pmax$ and $\rhomax$ and a lower limit of $\Rmax$ for a given $\Mmax$ (parametrized by $\rho_{\rm th}$). Excluding $\rho_{\rm th}$, we get $\Rmax > 1.41 r_{g*}$ [the widely-used constraint, the darkened area in Fig.~\ref{fig:MRPrho}(a)] and $\Pmax < 32.9\rho_0 c^2 (\text{\Msun}/\Mmax)^2$. Applying the latter limit\footnote{
In Ref.\ \cite{LattPrak2016}, the coefficient is $34.1$ since the authors use $\rho_0 = 2.7\times 10^{14}\,$g$\,$cm$^{-3}$. The maximum compactness condition also gives the relation between $\rhomax$ and $\Mmax$, but it appears less restrictive in the ranges shown in Fig.~\ref{fig:MRPrho}.
} to fits~(\ref{eq:x-MR}) and~(\ref{eq:y-Prho}), I obtain new constraints on properties of maximum-mass NSs. Upper boundaries of these constraints satisfy the equations
\begin{equation}
\label{eq:maxcomp-MR}
\Pmax(\Mmax,\Rmax)\times(1\pm 0.2) = 32.9 \rho_0 c^2 \left( \text{\Msun}/\Mmax \right)^2, 
\end{equation}
\begin{equation}
\label{eq:maxcomp-Prho}
\Mmax(\Pmax,\rhomax)\times(1\pm 0.03) = \left( 32.9\rho_0 c^2/\Pmax \right)^{1/2}.
\end{equation}
Again, the factors $(1\pm \ldots)$ account for fit errors. The solutions to these equations are shown in Fig.~\ref{fig:MRPrho} by long-dashed lines. They are somewhat less stringent than the direct causality conditions with the upper boundaries (\ref{eq:lumin-Rrg}) and~(\ref{eq:lumin-Prho}). However, they are still more restrictive than the direct maximum compactness $\Mmax-\Rmax$ condition. The boundaries of the observational constraints $1.97\,\text{\Msun} < \Mmax < 2.33\,\text{\Msun}$ are shown in Fig.~\ref{fig:MRPrho} by the short-dashed and dot-dashed lines. To plot them in the $\Pmax-\rhomax$ plane, one should use fit~(\ref{eq:y-Prho}) for $\Mmax$. It yields the $\Pmax-\rhomax$ relation for a fixed maximum mass,
\begin{multline}
\label{eq:Mfixed-Prho}
\left( \rho_0 c^2/\Pmax \right)^{1/5.32} - 2.56\left( \rho_0/\rhomax \right)^{1/1.72} 
\\
\approx 0.587 - 0.256\Mmax/\text{\Msun}\times(1\pm 0.03),
\end{multline}
where the factor $1\pm 0.03$ accounts for the errors. 

Another constraint was reported by Bauswein et al.~\cite{Baus2017}, who based it on the evidence that there was no direct black hole formation in the GW170817 event, and the merger remnant was stable at least for 10~ms. This restricts the binary mass threshold for a prompt collapse, $M_\text{thres} > M_\text{bin} + 0.1\,$\Msun, where $M_\text{bin} = 2.74_{-0.01}^{+0.04}\,$\Msun\ is the initial mass of this merged binary \cite{GW170817}. Then the authors of Ref.~\cite{Baus2017} used an EoS-independent approximate relation between $M_\text{thres}$, $\Mmax$, and $\Rmax$ to get some bounds on the latter quantity. Below we will use their formula with the lower credible limit for $M_\text{bin}$,
\begin{equation}
\label{eq:Bauswein}
\left( 2.43 - 1.69 r_{g*}/ \Rmax \right) \Mmax > 2.83\,\text{\Msun}.
\end{equation}
In Fig.~\ref{fig:MRPrho}(a), a lower boundary of this condition is shown by the double-dot-dashed line. It is mapped point-by-point to the $\Pmax-\rhomax$ plane [the double-dot-dashed line in Fig.~\ref{fig:MRPrho}(b)] using fits~(\ref{eq:x-MR}), with $\Pmax$ and $\rhomax$ uncertainties estimated as 20\% and 7\%, respectively.

\section{Discussion and Conclusion}

Eqs.~(\ref{eq:x-MR}) and~(\ref{eq:y-Prho}) fit the correlations between the quantities $\Mmax$, $\Rmax$, $\Pmax$, $\rhomax$, and $\csmax$ for various models of the maximum-mass NSs. These fits are calibrated for a particular set of EoSs which is wide enough to validate the suggestion that the correlations really exist, at least for the baryonic NS models. Their existence indicates that the $P-\rho$ relations for the chosen EoSs approximately form a family of 2-parameter functions. Similar result were obtained by Lindblom~\cite{Lind2010} for another set of EoSs, which partially overlaps with the collection used here. Thus, I assume that the $P-\rho$ relations of my EoS selection and of the Lindblom's one belong to the same 2-parameter family, which I find possible to call as ``the Lindblom 2-parameter family''.

Eqs.~(\ref{eq:x-MR}) and~(\ref{eq:y-Prho}) allow one to calculate any quantity --- $\Mmax$, $\Rmax$, $\Pmax$, $\rhomax$, or $\csmax$, if any other two are already known. In particular, these formulas give a point-by-point correspondence between the planes $\Mmax-\Rmax$ and $\Pmax-\rhomax$. If one has some constraint on the properties of the maximum-mass NSs, relating two of these five quantities, one can immediately translate this constraint to the other quantities. This gives a new method to explore the properties of the maximum-mass NSs. I have applied this method to several known constraints on the maximum-mass NSs and showed that it indeed allows one to derive new constraints on the properties of such stars.

Let us clarify physical meaning of these constraints focusing, for instance, on Eq.~(\ref{eq:lumin-Rrg}). It implies that if an EoS belongs to the Lindblom 2-parameter family, its $(\Mmax,\Rmax)$ point in Fig.~\ref{fig:MRPrho}(a) should be below the boundary given by Eq.~(\ref{eq:lumin-Rrg}). This ``if'' is very important. One should bear in mind that the correlations~(\ref{eq:x-MR}), which are basic for the constraint~(\ref{eq:lumin-Rrg}), reflect the properties of many theoretical EoSs. Although the EoS collection is large and representative, its properties are not a physical law. There can be other EoS models violating the correlations, and the real EoS of NS matter does not necessarily obey these correlations. The same ``if'' concerns all other constraints derived using fits~(\ref{eq:x-MR}) or~(\ref{eq:y-Prho}). If further investigations of NS EoSs indicate that  correlations~(\ref{eq:x-MR}) and~(\ref{eq:y-Prho}) are violated, it will suggest that the real EoS is different from the Lindblom family.

Moreover, among fits~(\ref{eq:x-MR}), the $\csmax(\Mmax,\Rmax)$ fit is the least reliable (see above). Hence, one may expect the causality condition to be the least reliable among all the restrictions plotted in Fig.~\ref{fig:MRPrho}. However, if $P(\rho)$ significantly deviates from the Lindblom 2-parameter 
family and violates the correlations, the EoS will likely be softer than stiffer. Then the solid lines in Fig.~\ref{fig:MRPrho} give at least an upper estimate of the ``true'' causality condition, and the restrictions that use this dependence remain valid.

If, however, the causality constraint obtained with the $\csmax(\Mmax,\Rmax)$ fit is accepted, it gives a novel restriction on the possible $M-R$ range for NSs (not only for the maximum-mass ones). The area on the $M-R$ plane above the solid line (with account for the uncertainties) should be excluded from this range. This would extend the domain forbidden by general relativity and causality, making it wider than the common condition $R<1.41\times 2GM/c^2$ described above.

Combining all the conditions in Fig.~\ref{fig:MRPrho} together gives a tight constraint on the domains where the points $(\Mmax,\Rmax)$ and $(\Pmax,\rhomax)$ can reside, if an EoS belongs to the Lindblom 2-parameter family (the hatched areas between the solid, short-dashed, dash-dotted, and double-dot-dashed lines in Fig.~\ref{fig:MRPrho}). In particular, this yields $\Rmax \gtrsim 9.6\,$km (similar to Ref.\ \cite{Baus2017}, which employed Eq.~(\ref{eq:Bauswein}) only), $\rhomax \lesssim 12\rho_0$, and $\Pmax \lesssim 7\rho_0 c^2$. 

Future observations of high-mass pulsars and NS-NS mergers may give new constraints on $\Mmax$ and set stronger constraints on $M_\text{thres}$, increasing the right-hand side of the condition~(\ref{eq:Bauswein}). In addition, one may prefer to consider the already-existing constraints on $\Mmax$ as more restrictive (e.g., $2.05\,\text{\Msun} \leqslant \Mmax \leqslant 2.17\,\text{\Msun}$, as it could be deduced from Refs.~\cite{Crom2019,MargMetz2017}). Then the above analysis would yield more strict constraints on $\Pmax$ and $\rhomax$.

Projecting the hatched area from Fig.~\ref{fig:MRPrho}(b) to the plane where full $P(\rho)$ curves are plotted, one can use this area to constrain the NS EoS. Namely, the EoS curve has to pass through this domain (and the segment of the curve after passing the domain is not actual for stable NSs).

Furthermore, let us compare the solid and short-dashed lines from Fig.~\ref{fig:MRPrho}(b) (causality and high-enough-mass conditions) with the $P(\rho)$ curves for the selected EoSs. Apparently, if an EoS has the $(\Pmax,\rhomax)$ point between these two lines (as plotted by a black filled symbol in Fig.~\ref{fig:MRPrho}), then the $P(\rho)$ curve of this EoS intersects neither the solid nor the short-dashed line at $3\rho_0\lesssim\rho<\rhomax$. Having no strict proof, I consider this fact as a viable assumption. Then the area between the solid and dashed lines forms the ($P,\rho$) domain that fundamentally can be reached in NSs at $\rho>3\rho_0$. 

Although no exotic phases of matter are allowed in the selected 50 EoSs, such exotic EoSs belong usually to the Lindblom 2-parameter family within the same range of errors in the $P-\rho$ space. Thus I expect that the proposed fits are good for describing properties of NSs containing exotic matter (except, maybe, those for $\csmax$). This allows one to speculate that the inferred limits on $\rho$ and $P$ are the density and pressure values, fundamentally reachable in any  stable stellar object. 

There are still many problems to be solved. One of them is the two-parameter nature of the EoS set. Another concern is that the object of this study is not a natural physical phenomenon, but a diverse family of theoretical EoSs. However, the most modern models in this family are widely expected to have $P(\rho)$ dependence reasonably close to that of the real EoS. If so, the presented results are also realistic.

\begin{acknowledgments}
I am grateful to P. Haensel and P.S.~Shternin for providing tables of some EoSs and valuable discussions as well as to M.E.~Gusakov and D.G.~Yakovlev for discussions. I also thank the anonymous referees for critical comments. The work was supported in part by the Foundation for the Advancement of Theoretical Physics and Mathematics ``BASIS'' (Grant No. 17-15-509-1) and by the Russian Foundation for Basic Research, project 19-52-12013 NNIO\_a. 
\end{acknowledgments}


\begin{thebibliography}{32}
\expandafter\ifx\csname natexlab\endcsname\relax\def\natexlab#1{#1}\fi
\expandafter\ifx\csname bibnamefont\endcsname\relax
  \def\bibnamefont#1{#1}\fi
\expandafter\ifx\csname bibfnamefont\endcsname\relax
  \def\bibfnamefont#1{#1}\fi
\expandafter\ifx\csname citenamefont\endcsname\relax
  \def\citenamefont#1{#1}\fi
\expandafter\ifx\csname url\endcsname\relax
  \def\url#1{\texttt{#1}}\fi
\expandafter\ifx\csname urlprefix\endcsname\relax\def\urlprefix{URL }\fi
\providecommand{\bibinfo}[2]{#2}
\providecommand{\eprint}[2][]{\url{#2}}

\bibitem[{\citenamefont{{Lattimer} and {Prakash}}(2016)}]{LattPrak2016}
\bibinfo{author}{\bibfnamefont{J.~M.} \bibnamefont{{Lattimer}}}
  \bibnamefont{and}
  \bibinfo{author}{\bibfnamefont{M.}~\bibnamefont{{Prakash}}},
  \bibinfo{journal}{\physrep} \textbf{\bibinfo{volume}{621}},
  \bibinfo{pages}{127} (\bibinfo{year}{2016}), \eprint{1512.07820}.

\bibitem[{\citenamefont{{Tolman}}(1939)}]{Tolman1939}
\bibinfo{author}{\bibfnamefont{R.~C.} \bibnamefont{{Tolman}}},
  \bibinfo{journal}{Physical Review} \textbf{\bibinfo{volume}{55}},
  \bibinfo{pages}{364} (\bibinfo{year}{1939}).

\bibitem[{\citenamefont{{Oppenheimer} and {Volkoff}}(1939)}]{OppVol1939}
\bibinfo{author}{\bibfnamefont{J.~R.} \bibnamefont{{Oppenheimer}}}
  \bibnamefont{and} \bibinfo{author}{\bibfnamefont{G.~M.}
  \bibnamefont{{Volkoff}}}, \bibinfo{journal}{Physical Review}
  \textbf{\bibinfo{volume}{55}}, \bibinfo{pages}{374} (\bibinfo{year}{1939}).

\bibitem[{\citenamefont{{Lindblom}}(1992)}]{Lind1992}
\bibinfo{author}{\bibfnamefont{L.}~\bibnamefont{{Lindblom}}},
  \bibinfo{journal}{\apj} \textbf{\bibinfo{volume}{398}}, \bibinfo{pages}{569}
  (\bibinfo{year}{1992}).

\bibitem[{\citenamefont{{Read} et~al.}(2009)\citenamefont{{Read}, {Lackey},
  {Owen}, and {Friedman}}}]{Read+2009}
\bibinfo{author}{\bibfnamefont{J.~S.} \bibnamefont{{Read}}},
  \bibinfo{author}{\bibfnamefont{B.~D.} \bibnamefont{{Lackey}}},
  \bibinfo{author}{\bibfnamefont{B.~J.} \bibnamefont{{Owen}}},
  \bibnamefont{and} \bibinfo{author}{\bibfnamefont{J.~L.}
  \bibnamefont{{Friedman}}}, \bibinfo{journal}{\prd}
  \textbf{\bibinfo{volume}{79}}, \bibinfo{eid}{124032} (\bibinfo{year}{2009}),
  \eprint{0812.2163}.

\bibitem[{\citenamefont{{Haensel} et~al.}(2007)\citenamefont{{Haensel},
  {Potekhin}, and {Yakovlev}}}]{HPY2007}
\bibinfo{author}{\bibfnamefont{P.}~\bibnamefont{{Haensel}}},
  \bibinfo{author}{\bibfnamefont{A.~Y.} \bibnamefont{{Potekhin}}},
  \bibnamefont{and} \bibinfo{author}{\bibfnamefont{D.~G.}
  \bibnamefont{{Yakovlev}}}, \emph{\bibinfo{title}{{Neutron Stars 1 : Equation
  of State and Structure}}}, vol. \bibinfo{volume}{326}
  (\bibinfo{publisher}{{Springer}}, \bibinfo{address}{{New York}},
  \bibinfo{year}{2007}).

\bibitem[{\citenamefont{{Antoniadis} et~al.}(2013)\citenamefont{{Antoniadis},
  {Freire}, {Wex}, {Tauris}, {Lynch}, {van Kerkwijk}, {Kramer}, {Bassa},
  {Dhillon}, {Driebe} et~al.}}]{Ant2013}
\bibinfo{author}{\bibfnamefont{J.}~\bibnamefont{{Antoniadis}}},
  \bibinfo{author}{\bibfnamefont{P.~C.~C.} \bibnamefont{{Freire}}},
  \bibinfo{author}{\bibfnamefont{N.}~\bibnamefont{{Wex}}},
  \bibinfo{author}{\bibfnamefont{T.~M.} \bibnamefont{{Tauris}}},
  \bibinfo{author}{\bibfnamefont{R.~S.} \bibnamefont{{Lynch}}},
  \bibinfo{author}{\bibfnamefont{M.~H.} \bibnamefont{{van Kerkwijk}}},
  \bibinfo{author}{\bibfnamefont{M.}~\bibnamefont{{Kramer}}},
  \bibinfo{author}{\bibfnamefont{C.}~\bibnamefont{{Bassa}}},
  \bibinfo{author}{\bibfnamefont{V.~S.} \bibnamefont{{Dhillon}}},
  \bibinfo{author}{\bibfnamefont{T.}~\bibnamefont{{Driebe}}},
  \bibnamefont{et~al.}, \bibinfo{journal}{Science}
  \textbf{\bibinfo{volume}{340}}, \bibinfo{pages}{448} (\bibinfo{year}{2013}),
  \eprint{1304.6875}.

\bibitem[{\citenamefont{{Cromartie} et~al.}(2019)\citenamefont{{Cromartie},
  {Fonseca}, {Ransom}, {Demorest}, {Arzoumanian}, {Blumer}, {Brook}, {DeCesar},
  {Dolch}, {Ellis} et~al.}}]{Crom2019}
\bibinfo{author}{\bibfnamefont{H.~T.} \bibnamefont{{Cromartie}}},
  \bibinfo{author}{\bibfnamefont{E.}~\bibnamefont{{Fonseca}}},
  \bibinfo{author}{\bibfnamefont{S.~M.} \bibnamefont{{Ransom}}},
  \bibinfo{author}{\bibfnamefont{P.~B.} \bibnamefont{{Demorest}}},
  \bibinfo{author}{\bibfnamefont{Z.}~\bibnamefont{{Arzoumanian}}},
  \bibinfo{author}{\bibfnamefont{H.}~\bibnamefont{{Blumer}}},
  \bibinfo{author}{\bibfnamefont{P.~R.} \bibnamefont{{Brook}}},
  \bibinfo{author}{\bibfnamefont{M.~E.} \bibnamefont{{DeCesar}}},
  \bibinfo{author}{\bibfnamefont{T.}~\bibnamefont{{Dolch}}},
  \bibinfo{author}{\bibfnamefont{J.~A.} \bibnamefont{{Ellis}}},
  \bibnamefont{et~al.}, \bibinfo{journal}{Nature Astronomy} p.
  \bibinfo{pages}{439} (\bibinfo{year}{2019}), \eprint{1904.06759}.

\bibitem[{\citenamefont{{Abbott} et~al.}(2017)\citenamefont{{Abbott}, {Abbott},
  {Abbott} et~al.}}]{GW170817}
\bibinfo{author}{\bibfnamefont{B.~P.} \bibnamefont{{Abbott}}},
  \bibinfo{author}{\bibfnamefont{R.}~\bibnamefont{{Abbott}}},
  \bibinfo{author}{\bibfnamefont{T.~D.} \bibnamefont{{Abbott}}},
  \bibnamefont{et~al.}, \bibinfo{journal}{\prl} \textbf{\bibinfo{volume}{119}},
  \bibinfo{eid}{161101} (\bibinfo{year}{2017}), \eprint{1710.05832}.

\bibitem[{\citenamefont{{Rezzolla} et~al.}(2018)\citenamefont{{Rezzolla},
  {Most}, and {Weih}}}]{Rezz2018}
\bibinfo{author}{\bibfnamefont{L.}~\bibnamefont{{Rezzolla}}},
  \bibinfo{author}{\bibfnamefont{E.~R.} \bibnamefont{{Most}}},
  \bibnamefont{and} \bibinfo{author}{\bibfnamefont{L.~R.}
  \bibnamefont{{Weih}}}, \bibinfo{journal}{\apjl}
  \textbf{\bibinfo{volume}{852}}, \bibinfo{eid}{L25} (\bibinfo{year}{2018}),
  \eprint{1711.00314}.

\bibitem[{\citenamefont{{Fiorella Burgio} and {Fantina}}(2018)}]{BurgFant2018}
\bibinfo{author}{\bibfnamefont{G.}~\bibnamefont{{Fiorella Burgio}}}
  \bibnamefont{and} \bibinfo{author}{\bibfnamefont{A.~F.}
  \bibnamefont{{Fantina}}}, in \emph{\bibinfo{booktitle}{The Physics and
  Astrophysics of Neutron Stars}}, edited by
  \bibinfo{editor}{\bibfnamefont{L.}~\bibnamefont{{Rezzolla}}},
  \bibinfo{editor}{\bibfnamefont{P.}~\bibnamefont{{Pizzochero}}},
  \bibinfo{editor}{\bibfnamefont{D.~I.} \bibnamefont{{Jones}}},
  \bibinfo{editor}{\bibfnamefont{N.}~\bibnamefont{{Rea}}}, \bibnamefont{and}
  \bibinfo{editor}{\bibfnamefont{I.}~\bibnamefont{{Vida{\~n}a}}}, 
  vol. \bibinfo{volume}{457} of
  \emph{\bibinfo{series}{Astrophysics and Space Science Library}} 
  (\bibinfo{publisher}{Springer}, \bibinfo{address}{Cham}, 
  \bibinfo{year}{2018}), p. \bibinfo{pages}{255}, \eprint{1804.03020}.

\bibitem[{\citenamefont{{Lindblom}}(2010)}]{Lind2010}
\bibinfo{author}{\bibfnamefont{L.}~\bibnamefont{{Lindblom}}},
  \bibinfo{journal}{\prd} \textbf{\bibinfo{volume}{82}}, \bibinfo{eid}{103011}
  (\bibinfo{year}{2010}), \eprint{1009.0738}.

\bibitem[{\citenamefont{{Prakash} et~al.}(1988)\citenamefont{{Prakash},
  {Ainsworth}, and {Lattimer}}}]{PAL1988}
\bibinfo{author}{\bibfnamefont{M.}~\bibnamefont{{Prakash}}},
  \bibinfo{author}{\bibfnamefont{T.~L.} \bibnamefont{{Ainsworth}}},
  \bibnamefont{and} \bibinfo{author}{\bibfnamefont{J.~M.}
  \bibnamefont{{Lattimer}}}, \bibinfo{journal}{\prl}
  \textbf{\bibinfo{volume}{61}}, \bibinfo{pages}{2518} (\bibinfo{year}{1988}).

\bibitem[{\citenamefont{{Page} and {Applegate}}(1992)}]{PAPAL1992}
\bibinfo{author}{\bibfnamefont{D.}~\bibnamefont{{Page}}} \bibnamefont{and}
  \bibinfo{author}{\bibfnamefont{J.~H.} \bibnamefont{{Applegate}}},
  \bibinfo{journal}{\apjl} \textbf{\bibinfo{volume}{394}}, \bibinfo{pages}{L17}
  (\bibinfo{year}{1992}).

\bibitem[{\citenamefont{{Yakovlev} et~al.}(2011)\citenamefont{{Yakovlev}, {Ho},
  {Shternin}, {Heinke}, and {Potekhin}}}]{Yak2011}
\bibinfo{author}{\bibfnamefont{D.~G.} \bibnamefont{{Yakovlev}}},
  \bibinfo{author}{\bibfnamefont{W.~C.~G.} \bibnamefont{{Ho}}},
  \bibinfo{author}{\bibfnamefont{P.~S.} \bibnamefont{{Shternin}}},
  \bibinfo{author}{\bibfnamefont{C.~O.} \bibnamefont{{Heinke}}},
  \bibnamefont{and} \bibinfo{author}{\bibfnamefont{A.~Y.}
  \bibnamefont{{Potekhin}}}, \bibinfo{journal}{\mnras}
  \textbf{\bibinfo{volume}{411}}, \bibinfo{pages}{1977} (\bibinfo{year}{2011}),
  \eprint{1010.1154}.

\bibitem[{\citenamefont{{Akmal} et~al.}(1998)\citenamefont{{Akmal},
  {Pandharipande}, and {Ravenhall}}}]{APR1998}
\bibinfo{author}{\bibfnamefont{A.}~\bibnamefont{{Akmal}}},
  \bibinfo{author}{\bibfnamefont{V.~R.} \bibnamefont{{Pandharipande}}},
  \bibnamefont{and} \bibinfo{author}{\bibfnamefont{D.~G.}
  \bibnamefont{{Ravenhall}}}, \bibinfo{journal}{\prc}
  \textbf{\bibinfo{volume}{58}}, \bibinfo{pages}{1804} (\bibinfo{year}{1998}),
  \eprint{nucl-th/9804027}.

\bibitem[{\citenamefont{{Heiselberg} and {Hjorth-Jensen}}(2000)}]{HHJ2000}
\bibinfo{author}{\bibfnamefont{H.}~\bibnamefont{{Heiselberg}}}
  \bibnamefont{and}
  \bibinfo{author}{\bibfnamefont{M.}~\bibnamefont{{Hjorth-Jensen}}},
  \bibinfo{journal}{\physrep} \textbf{\bibinfo{volume}{328}},
  \bibinfo{pages}{237} (\bibinfo{year}{2000}), \eprint{nucl-th/9902033}.

\bibitem[{\citenamefont{{Gusakov} et~al.}(2005)\citenamefont{{Gusakov},
  {Kaminker}, {Yakovlev}, and {Gnedin}}}]{Gus2005}
\bibinfo{author}{\bibfnamefont{M.~E.} \bibnamefont{{Gusakov}}},
  \bibinfo{author}{\bibfnamefont{A.~D.} \bibnamefont{{Kaminker}}},
  \bibinfo{author}{\bibfnamefont{D.~G.} \bibnamefont{{Yakovlev}}},
  \bibnamefont{and} \bibinfo{author}{\bibfnamefont{O.~Y.}
  \bibnamefont{{Gnedin}}}, \bibinfo{journal}{\mnras}
  \textbf{\bibinfo{volume}{363}}, \bibinfo{pages}{555} (\bibinfo{year}{2005}),
  \eprint{astro-ph/0507560}.

\bibitem[{\citenamefont{{Kaminker} et~al.}(2014)\citenamefont{{Kaminker},
  {Kaurov}, {Potekhin}, and {Yakovlev}}}]{KKPY2014}
\bibinfo{author}{\bibfnamefont{A.~D.} \bibnamefont{{Kaminker}}},
  \bibinfo{author}{\bibfnamefont{A.~A.} \bibnamefont{{Kaurov}}},
  \bibinfo{author}{\bibfnamefont{A.~Y.} \bibnamefont{{Potekhin}}},
  \bibnamefont{and} \bibinfo{author}{\bibfnamefont{D.~G.}
  \bibnamefont{{Yakovlev}}}, \bibinfo{journal}{\mnras}
  \textbf{\bibinfo{volume}{442}}, \bibinfo{pages}{3484} (\bibinfo{year}{2014}),
  \eprint{1406.0723}.

\bibitem[{\citenamefont{{Douchin} and {Haensel}}(2001)}]{DH2001}
\bibinfo{author}{\bibfnamefont{F.}~\bibnamefont{{Douchin}}} \bibnamefont{and}
  \bibinfo{author}{\bibfnamefont{P.}~\bibnamefont{{Haensel}}},
  \bibinfo{journal}{\aap} \textbf{\bibinfo{volume}{380}}, \bibinfo{pages}{151}
  (\bibinfo{year}{2001}), \eprint{astro-ph/0111092}.

\bibitem[{\citenamefont{{Potekhin} et~al.}(2013)\citenamefont{{Potekhin},
  {Fantina}, {Chamel}, {Pearson}, and {Goriely}}}]{BSk2013}
\bibinfo{author}{\bibfnamefont{A.~Y.} \bibnamefont{{Potekhin}}},
  \bibinfo{author}{\bibfnamefont{A.~F.} \bibnamefont{{Fantina}}},
  \bibinfo{author}{\bibfnamefont{N.}~\bibnamefont{{Chamel}}},
  \bibinfo{author}{\bibfnamefont{J.~M.} \bibnamefont{{Pearson}}},
  \bibnamefont{and}
  \bibinfo{author}{\bibfnamefont{S.}~\bibnamefont{{Goriely}}},
  \bibinfo{journal}{\aap} \textbf{\bibinfo{volume}{560}}, \bibinfo{eid}{A48}
  (\bibinfo{year}{2013}), \eprint{1310.0049}.

\bibitem[{\citenamefont{{Pearson} et~al.}(2018)\citenamefont{{Pearson},
  {Chamel}, {Potekhin}, {Fantina}, {Ducoin}, {Dutta}, and {Goriely}}}]{BSk2018}
\bibinfo{author}{\bibfnamefont{J.~M.} \bibnamefont{{Pearson}}},
  \bibinfo{author}{\bibfnamefont{N.}~\bibnamefont{{Chamel}}},
  \bibinfo{author}{\bibfnamefont{A.~Y.} \bibnamefont{{Potekhin}}},
  \bibinfo{author}{\bibfnamefont{A.~F.} \bibnamefont{{Fantina}}},
  \bibinfo{author}{\bibfnamefont{C.}~\bibnamefont{{Ducoin}}},
  \bibinfo{author}{\bibfnamefont{A.~K.} \bibnamefont{{Dutta}}},
  \bibnamefont{and}
  \bibinfo{author}{\bibfnamefont{S.}~\bibnamefont{{Goriely}}},
  \bibinfo{journal}{\mnras} \textbf{\bibinfo{volume}{481}},
  \bibinfo{pages}{2994} (\bibinfo{year}{2018}), \eprint{1903.04981}.

\bibitem[{\citenamefont{{Shternin} et~al.}(2018)\citenamefont{{Shternin},
  {Baldo}, and {Haensel}}}]{SBH2018}
\bibinfo{author}{\bibfnamefont{P.~S.} \bibnamefont{{Shternin}}},
  \bibinfo{author}{\bibfnamefont{M.}~\bibnamefont{{Baldo}}}, \bibnamefont{and}
  \bibinfo{author}{\bibfnamefont{P.}~\bibnamefont{{Haensel}}},
  \bibinfo{journal}{Physics Letters B} \textbf{\bibinfo{volume}{786}},
  \bibinfo{pages}{28} (\bibinfo{year}{2018}), \eprint{1807.06569}.

\bibitem[{\citenamefont{{Fortin} et~al.}(2016)\citenamefont{{Fortin},
  {Provid{\^e}ncia}, {Raduta}, {Gulminelli}, {Zdunik}, {Haensel}, and
  {Bejger}}}]{Fortin2016}
\bibinfo{author}{\bibfnamefont{M.}~\bibnamefont{{Fortin}}},
  \bibinfo{author}{\bibfnamefont{C.}~\bibnamefont{{Provid{\^e}ncia}}},
  \bibinfo{author}{\bibfnamefont{A.~R.} \bibnamefont{{Raduta}}},
  \bibinfo{author}{\bibfnamefont{F.}~\bibnamefont{{Gulminelli}}},
  \bibinfo{author}{\bibfnamefont{J.~L.} \bibnamefont{{Zdunik}}},
  \bibinfo{author}{\bibfnamefont{P.}~\bibnamefont{{Haensel}}},
  \bibnamefont{and} \bibinfo{author}{\bibfnamefont{M.}~\bibnamefont{{Bejger}}},
  \bibinfo{journal}{\prc} \textbf{\bibinfo{volume}{94}}, \bibinfo{eid}{035804}
  (\bibinfo{year}{2016}), \eprint{1604.01944}.

\bibitem[{\citenamefont{{Gusakov} et~al.}(2014)\citenamefont{{Gusakov},
  {Haensel}, and {Kantor}}}]{GHK2014}
\bibinfo{author}{\bibfnamefont{M.~E.} \bibnamefont{{Gusakov}}},
  \bibinfo{author}{\bibfnamefont{P.}~\bibnamefont{{Haensel}}},
  \bibnamefont{and} \bibinfo{author}{\bibfnamefont{E.~M.}
  \bibnamefont{{Kantor}}}, \bibinfo{journal}{\mnras}
  \textbf{\bibinfo{volume}{439}}, \bibinfo{pages}{318} (\bibinfo{year}{2014}),
  \eprint{1401.2827}.

\bibitem[{\citenamefont{{Horowitz} and {Piekarewicz}}(2001)}]{Horowitz01}
\bibinfo{author}{\bibfnamefont{C.~J.} \bibnamefont{{Horowitz}}}
  \bibnamefont{and}
  \bibinfo{author}{\bibfnamefont{J.}~\bibnamefont{{Piekarewicz}}},
  \bibinfo{journal}{\prl} \textbf{\bibinfo{volume}{86}}, \bibinfo{pages}{5647}
  (\bibinfo{year}{2001}), \eprint{astro-ph/0010227}.

\bibitem[{\citenamefont{{Provid{\^e}ncia}
  et~al.}(2019)\citenamefont{{Provid{\^e}ncia}, {Fortin}, {Pais}, and
  {Rabhi}}}]{Prov+2018}
\bibinfo{author}{\bibfnamefont{C.}~\bibnamefont{{Provid{\^e}ncia}}},
  \bibinfo{author}{\bibfnamefont{M.}~\bibnamefont{{Fortin}}},
  \bibinfo{author}{\bibfnamefont{H.}~\bibnamefont{{Pais}}}, \bibnamefont{and}
  \bibinfo{author}{\bibfnamefont{A.}~\bibnamefont{{Rabhi}}},
  \bibinfo{journal}{Frontiers in Astronomy and Space Sciences}
  \textbf{\bibinfo{volume}{6}}, \bibinfo{eid}{13} (\bibinfo{year}{2019}),
  \eprint{1811.00786}.

\bibitem[{\citenamefont{{Shapiro} and {Teukolsky}}(1983)}]{ShapTeuk1983}
\bibinfo{author}{\bibfnamefont{S.~L.} \bibnamefont{{Shapiro}}}
  \bibnamefont{and} \bibinfo{author}{\bibfnamefont{S.~A.}
  \bibnamefont{{Teukolsky}}}, \emph{\bibinfo{title}{{Black holes, white dwarfs,
  and neutron stars : the physics of compact objects}}} (\bibinfo{publisher}{{A
  Wiley-Interscience Publication}}, \bibinfo{address}{{New York}},
  \bibinfo{year}{1983}).

\bibitem[{\citenamefont{{Tews} et~al.}(2018)\citenamefont{{Tews}, {Carlson},
  {Gandolfi}, and {Reddy}}}]{Tews+2018}
\bibinfo{author}{\bibfnamefont{I.}~\bibnamefont{{Tews}}},
  \bibinfo{author}{\bibfnamefont{J.}~\bibnamefont{{Carlson}}},
  \bibinfo{author}{\bibfnamefont{S.}~\bibnamefont{{Gandolfi}}},
  \bibnamefont{and} \bibinfo{author}{\bibfnamefont{S.}~\bibnamefont{{Reddy}}},
  \bibinfo{journal}{\apj} \textbf{\bibinfo{volume}{860}}, \bibinfo{eid}{149}
  (\bibinfo{year}{2018}), \eprint{1801.01923}.

\bibitem[{\citenamefont{{Greif} et~al.}(2019)\citenamefont{{Greif},
  {Raaijmakers}, {Hebeler}, {Schwenk}, and {Watts}}}]{Greif+2019}
\bibinfo{author}{\bibfnamefont{S.~K.} \bibnamefont{{Greif}}},
  \bibinfo{author}{\bibfnamefont{G.}~\bibnamefont{{Raaijmakers}}},
  \bibinfo{author}{\bibfnamefont{K.}~\bibnamefont{{Hebeler}}},
  \bibinfo{author}{\bibfnamefont{A.}~\bibnamefont{{Schwenk}}},
  \bibnamefont{and} \bibinfo{author}{\bibfnamefont{A.~L.}
  \bibnamefont{{Watts}}}, \bibinfo{journal}{\mnras}
  \textbf{\bibinfo{volume}{485}}, \bibinfo{pages}{5363} (\bibinfo{year}{2019}),
  \eprint{1812.08188}.

\bibitem[{\citenamefont{{Bauswein} et~al.}(2017)\citenamefont{{Bauswein},
  {Just}, {Janka}, and {Stergioulas}}}]{Baus2017}
\bibinfo{author}{\bibfnamefont{A.}~\bibnamefont{{Bauswein}}},
  \bibinfo{author}{\bibfnamefont{O.}~\bibnamefont{{Just}}},
  \bibinfo{author}{\bibfnamefont{H.-T.} \bibnamefont{{Janka}}},
  \bibnamefont{and}
  \bibinfo{author}{\bibfnamefont{N.}~\bibnamefont{{Stergioulas}}},
  \bibinfo{journal}{\apjl} \textbf{\bibinfo{volume}{850}}, \bibinfo{eid}{L34}
  (\bibinfo{year}{2017}), \eprint{1710.06843}.

\bibitem[{\citenamefont{{Margalit} and {Metzger}}(2017)}]{MargMetz2017}
\bibinfo{author}{\bibfnamefont{B.}~\bibnamefont{{Margalit}}} \bibnamefont{and}
  \bibinfo{author}{\bibfnamefont{B.~D.} \bibnamefont{{Metzger}}},
  \bibinfo{journal}{\apjl} \textbf{\bibinfo{volume}{850}}, \bibinfo{eid}{L19}
  (\bibinfo{year}{2017}), \eprint{1710.05938}.

\end{thebibliography}


\end{document}